\documentclass[twocolumn,aps,pre,floatfix]{revtex4}
\usepackage{graphicx,color}
\usepackage{amsmath,amssymb}
\newcommand{\fignewx}{%
\begin{figure}
\includegraphics[width=0.45\textwidth,clip]{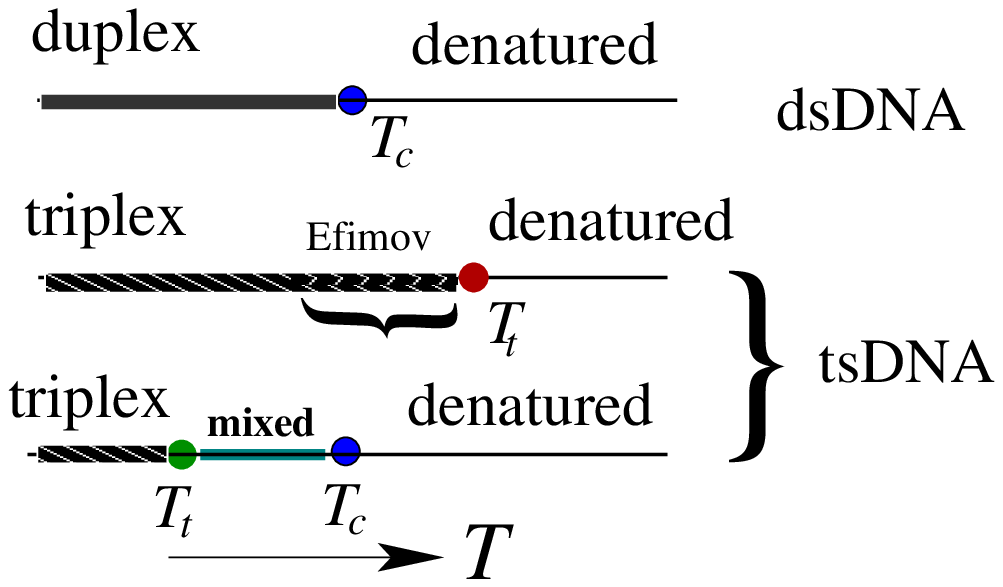}

\caption{Schematic diagram showing the predicted phases
  as temperature $T$ is varied.  While the Efimov-DNA is not a phase,
  mixed phase is predicted to be a thermodynamic phase.  A triplex DNA
  crosses over to the Efimov-DNA before melting at $T_t$, while in the
  case of a mixed phase, there will be a two-step melting, a
  triplex-to-mixed transition at $T_t$, and a mixed-to-denatured DNA
  transition at $T_c$.}
\label{fig:newx}
\end{figure}
}%

\newcommand{\cmodel}{%
\begin{figure}[htbp]
   \centering
   \includegraphics[width=.45\textwidth,clip]{JM_fig4a.eps}

\vspace{0.5cm}

   \includegraphics[width=.45\textwidth,clip]{JM_fig4b.eps}

   \caption{The two-chain phase diagram for $\sigma$ {\it vs} $y$.
     (a) The polymers can cross each other, (b) the noncrossing case.
     For both the cases, the two-chain melting is at $y_c(0)=1.264...$
     for $\sigma=0$. Here and elsewhere, $y=1$ ($y=\infty$) corresponds to
     infinite (zero) temperature.}

   \label{fig:cross}
 \end{figure}
}

\newcommand{\comodel}{%
\begin{figure}[htbp]
   \centering
   \includegraphics[width=0.45\textwidth]{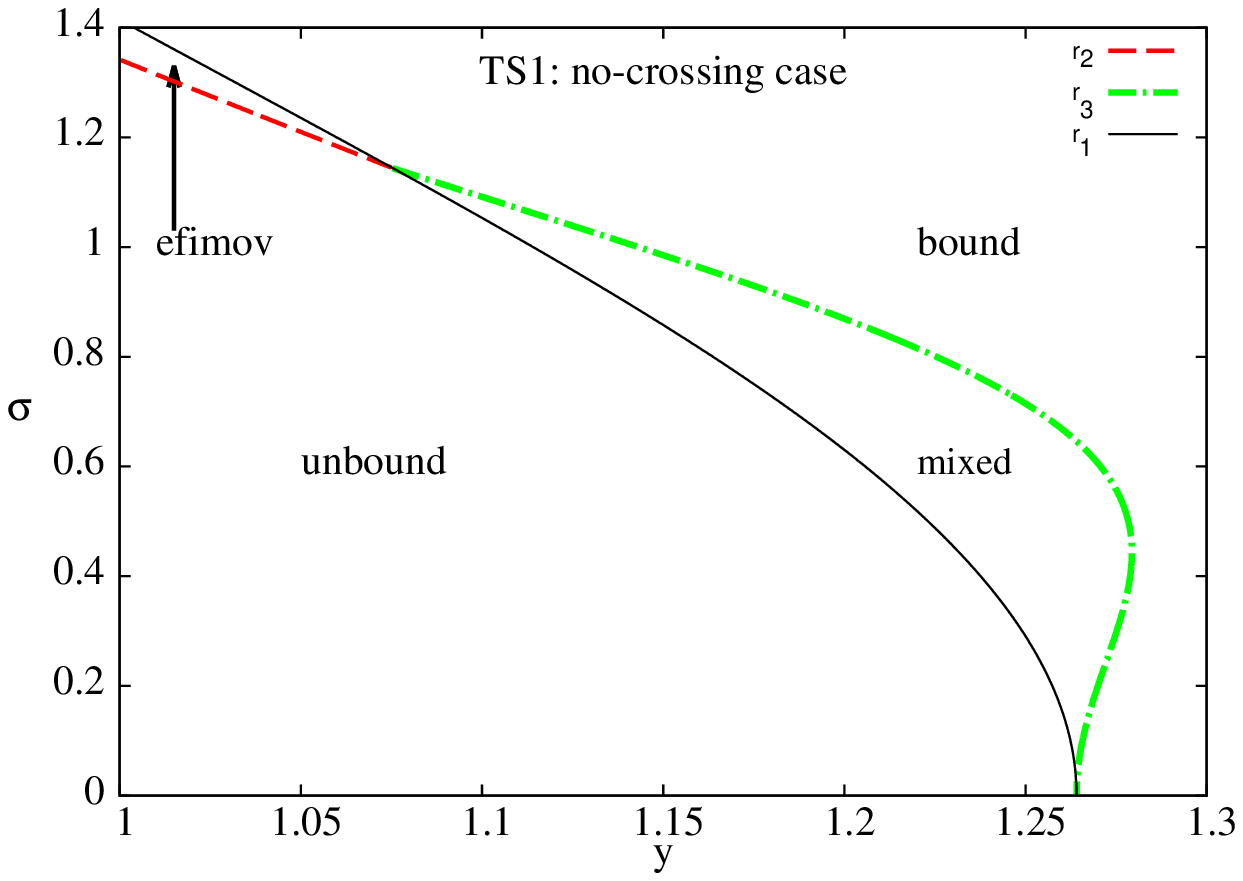}

   \caption{Model TS1: The three-chain phase diagram in
     the $y$--$\sigma$ plane for model TS1.  The bound, the unbound,
     the Efimov, and the mixed states are shown.  The solid line is the
     two-chain melting curve and is valid for the three chain case in
     the region $y>y_E=1.07526...$ but not in the region $y<y_E$.  The
     dashed-dotted (green) line is the boundary for the mixed phase while 
     the dashed line is the melting line for the Efimov-DNA. }

   \label{fig:s1}
 \end{figure}
}
\newcommand{\aomodeltwo}{%
\begin{figure}[htbp]
   \centering
   \includegraphics[width=.45\textwidth,clip]{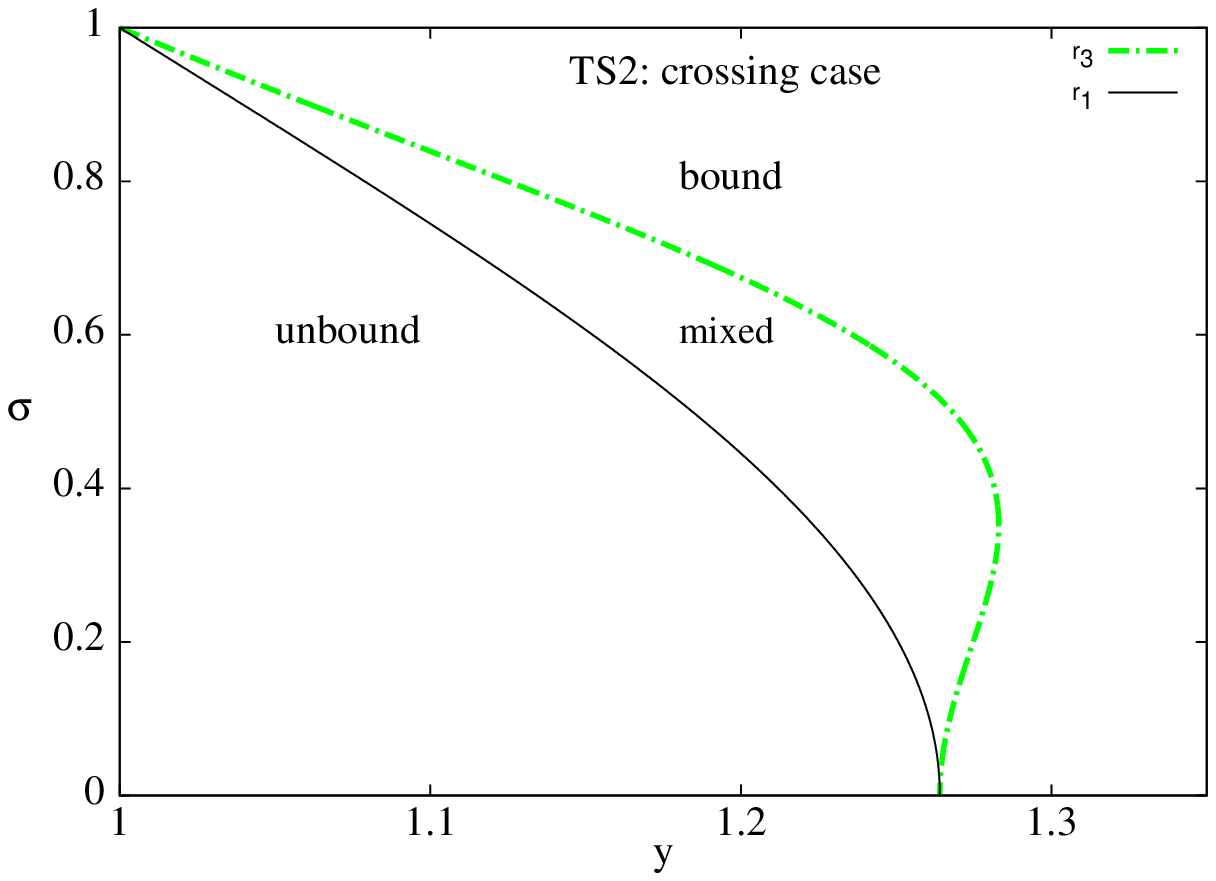}

   \caption{Model TS2: The three-chain phase diagram in
     the $y$-$\sigma$ plane.  The unbound, the bound, and the mixed
     phases are shown.  The solid line is the two-chain melting curve
     which is present in the three-chain case.  There is no Efimov-DNA.  
     The dash-dotted (green) line is the boundary for the mixed phase.  }

   \label{fig:s22}
 \end{figure}
}
\newcommand{\aomodelthree}{%
\begin{figure}[htbp]
   \centering
   \includegraphics[width=.43\textwidth,clip]{JM_fig8.eps}

   \caption{Model TS3: the three-chain phase diagram in
     the $y$-$\sigma$ plane.  The unbound, the bound, and the Efimov
     states are shown.  The dark (black) line representing the two-chain
     melting curve, is {\it not} present in the three-chain case.
     There is no mixed phase. }

   \label{fig:s23}
 \end{figure}
}

\newcommand{\senergy}{%
\begin{figure}[htbp]
   \centering
   \includegraphics[width=0.45\textwidth,clip]{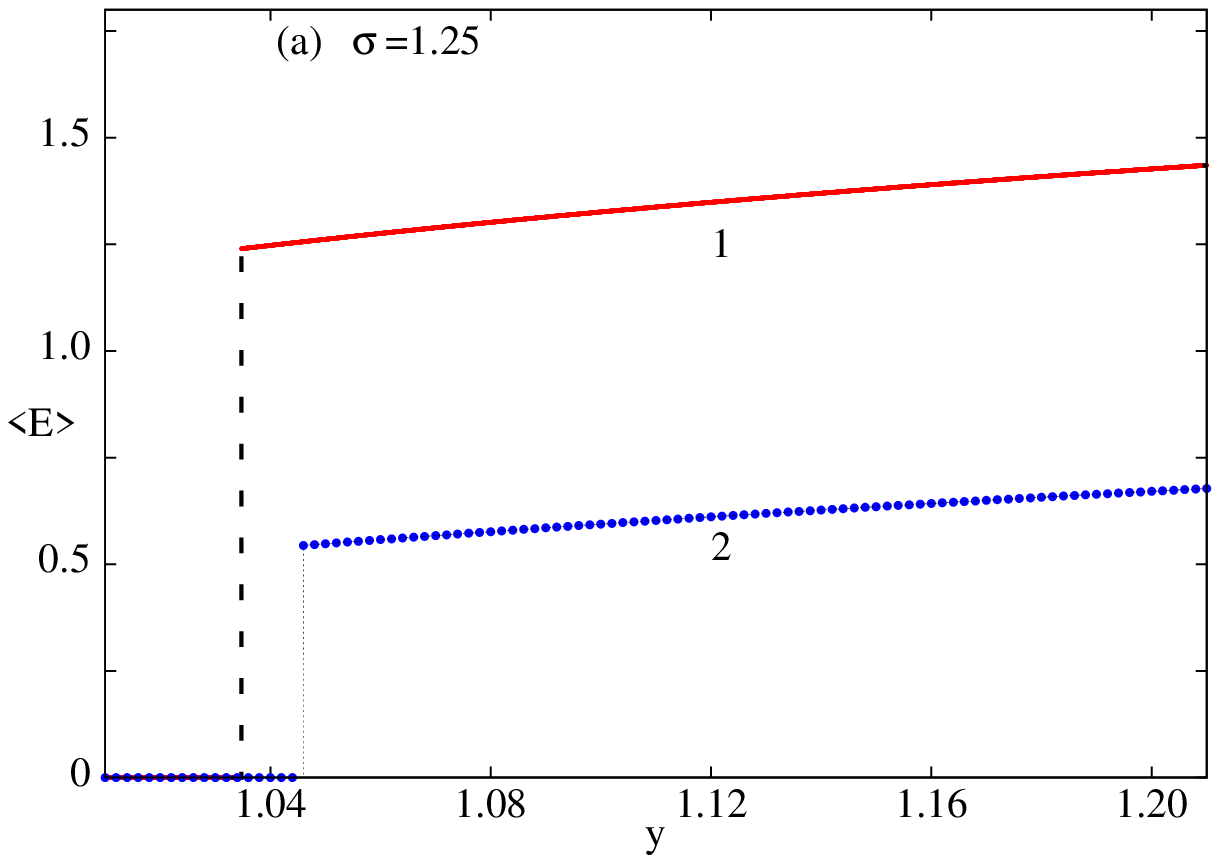}

   \includegraphics[width=0.45\textwidth,clip]{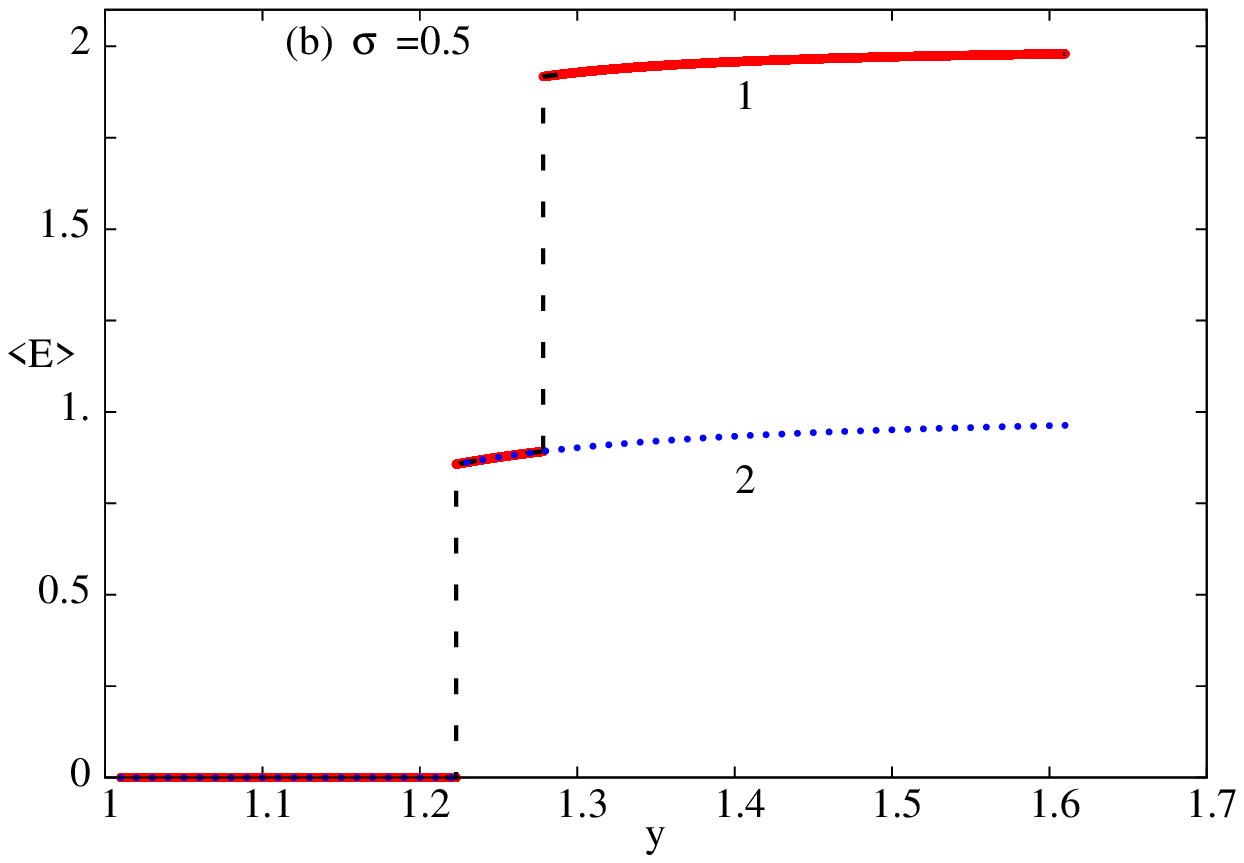}

   \caption{Plot of the average energy per bond with
     $y$, for Model TS1.  Here $\epsilon=1$.  The vertical lines 
     are to show the discontinuity in the energy curves. The three-chain
     average energy (marked as 1) is compared to the two-chain average
     energy (marked as 2).  (a) For $\sigma=1.25$.  (b) For
     $\sigma=0.5$.  The first order transition for the two chain model
     is consistent with experimental findings.}
   \label{fig:ss2}
 \end{figure}
}

\newcommand{\GSmodel}{%
\begin{figure}[htbp]
   \centering
   \includegraphics[width=0.45\textwidth]{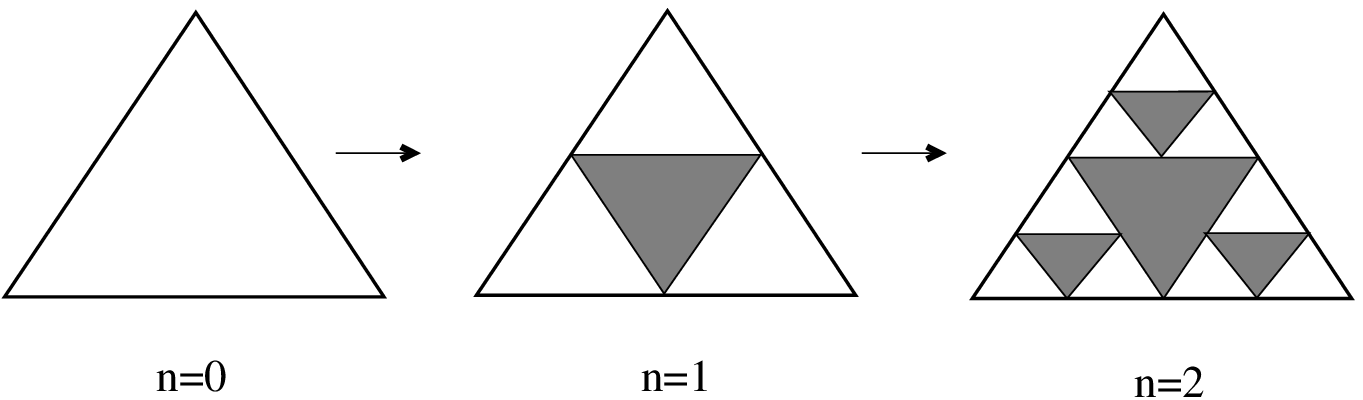}

   \caption{Recursive construction of the Sierpinski gasket.
}
   \label{fig:gsmodel}
 \end{figure}
}
\newcommand{\GSwalk}{%
\begin{figure}[htbp]
   \centering
   \includegraphics[width=0.45\textwidth]{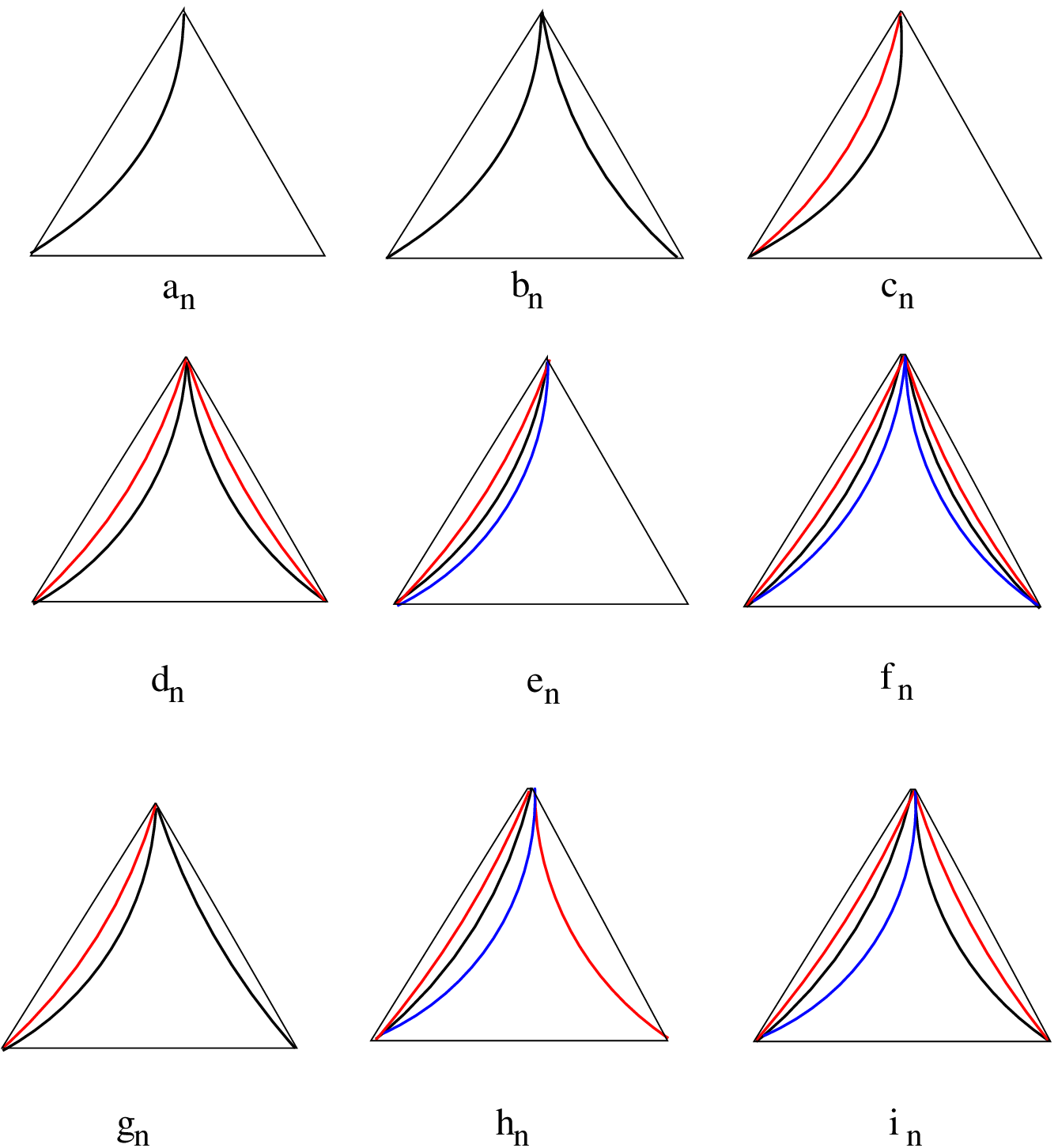}

   \caption{The partition functions for two and three
     strands irrespective of crossing conditions of the chains.  }
   \label{fig:gswalk}
 \end{figure}
}

\newcommand{\GSwalkM}{%
\begin{figure}[htbp]
   \centering
   \includegraphics[width=0.43\textwidth]{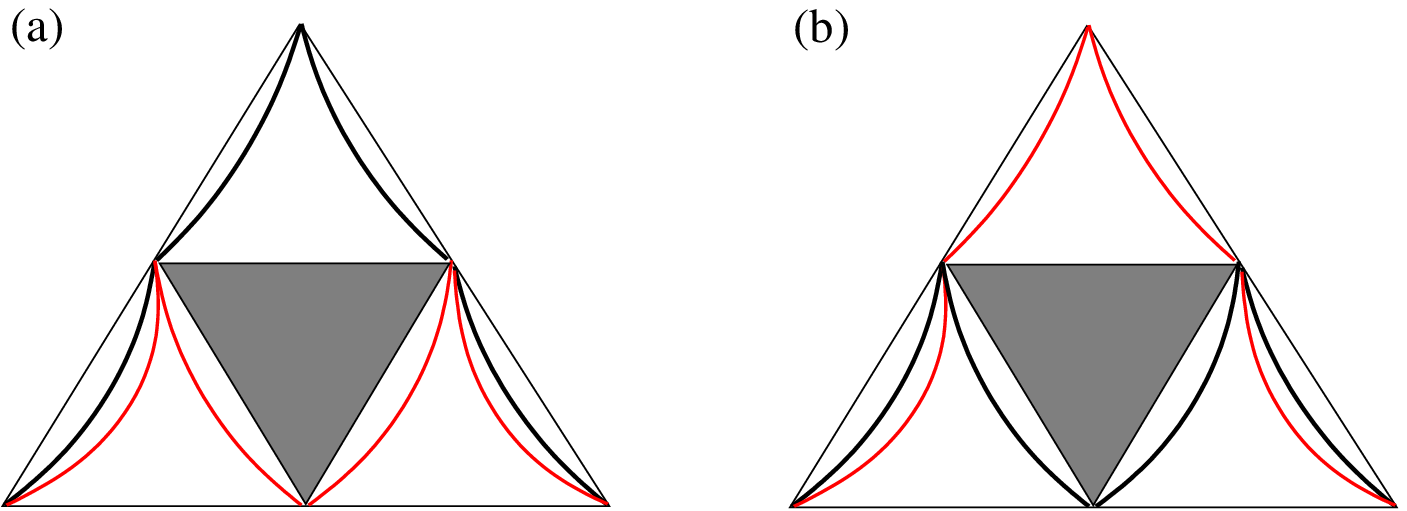}

   \caption{A polymer is not allowed on any horizontal
     bond.  In the figure, two possible configurations of two polymers
     (black (dark) and red (light)) are shown ($b_n$ and $g_n$ type
     from Fig.~\ref{fig:gswalk}).  The crossing case allows both while
     only (a) is allowed for the noncrossing case.  }
   \label{fig:gswalk1}
 \end{figure}
}

\newcommand{\fugac}{%
\begin{figure}[htbp]
   \centering
   \includegraphics[width=0.47\textwidth]{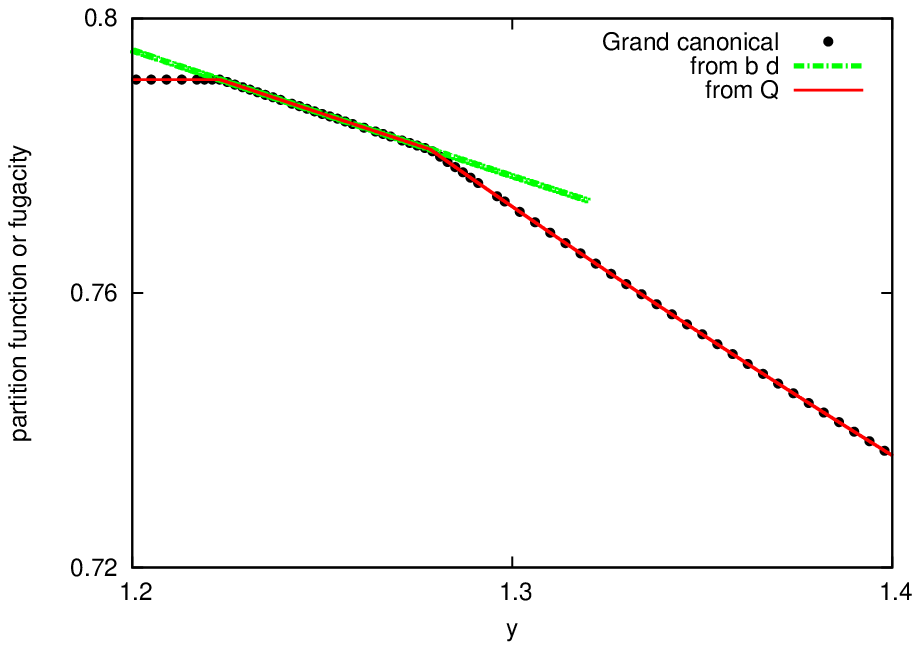}

   \caption{The fixed point values (circles) of grand
     canonical $z$ are compared with the canonical partition function
     $Q_{\rm tot}^{1/3N}$ (solid line). The values of the mixed phase
     partition function $(b_n d_{n})^{1/3N}$, are shown by the 
     dash-dotted (green) line.  Here $N=2^{26}$ is the length of each 
     polymer.}
   \label{fig:compar}
 \end{figure}
}

\newcommand{\mixed}{%
\begin{figure}[htbp]
   \centering
   \includegraphics[width=0.25\textwidth]{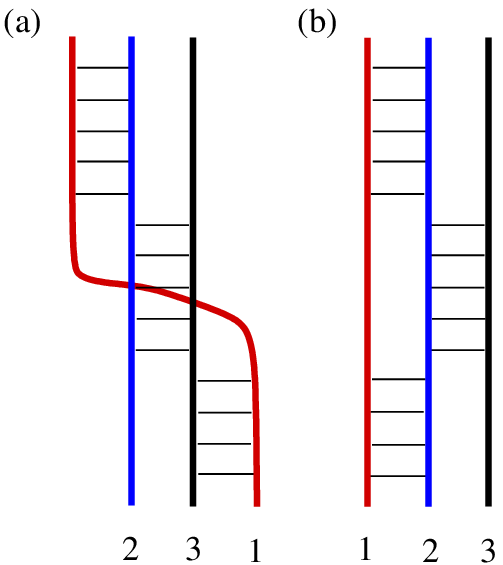}

   \caption{Schematic diagram of a mixed phase of three
     polymers of two possible configurations.  At each monomer
     position, two are bound but the third monomer is free along the
     length of the chains.  (a) Polymer chains can cross each other.
     (b) Polymer chains can not cross each other and no interaction
     between chains 1 and 3.  }
   \label{fig:mixed}
 \end{figure}
}

\begin{document}

\title{Melting behavior and different bound states in three stranded 
DNA models}

\author{Jaya Maji} \email{jayamaji@iopb.res.in} 
\author{Somendra M. Bhattacharjee} \email{somen@iopb.res.in} 
\affiliation{ Institute of  Physics, Bhubaneswar 751005, India} 
\author{Flavio Seno}\email{flavio.seno@pd.infn.it} 
\author{Antonio Trovato}\email{antonio.trovato@pd.infn.it} 
\affiliation{INFN, Dipartimento di Fisica e Astronomia `Galileo 
Galilei', Universit\`a di Padova, Via Marzolo 8, 35131 Padova, Italy}

\begin{abstract}
  Thermal denaturation of DNA is often studied with coarse-grained
  models in which native sequential base pairing is mimicked by the
  existence of attractive interactions only between monomers at the
  same position along strands (Poland and Scheraga models). Within
  this framework, the existence of a three strand DNA bound state in
  conditions where a duplex DNA would be in the denaturated state was
  recently predicted from a study of three directed polymer models on
  simplified hierarchical lattices ($d>2$) and in $1+1$ dimensions.
  Such phenomenon which is similar to the Efimov effect in nuclear
  physics was named Efimov-DNA. In this paper we study the melting of
  the three-stranded DNA on a Sierpinski gasket of dimensions $d<2$ by
  assigning extra weight factors to fork openings and closings, to
  induce a two-strand DNA melting.  In such a context we can find again
  the existence of the Efimov-DNA-like state but quite surprisingly we
  discover also the presence of a different phase, to be called a mixed
  state, where the strands are pair-wise bound but without three chain
  contacts.  Whereas the Efimov DNA turns out to be a crossover near
  melting, the mixed phase is a thermodynamic phase.
\end{abstract}
\date{\today}

\maketitle

\section{INTRODUCTION}
A loosely bound state of a triple stranded DNA when no two are bound
was recently found with a theoretical approach and named
Efimov-DNA \cite{efidna,zeros,tanlim}.  It occurs at and above the
melting point of a double stranded DNA
(dsDNA) \cite{melt,unzip1,unzip2,unzip3,kapri}, and is reminiscent of
the Efimov effect in quantum mechanics \cite{efimov1,efimov2}.  In
fact, the sequential base pairing of a DNA opens up a path to make a
formal connection between a quantum problem and the DNA
thermodynamics, with thermal fluctuations playing the role of quantum
fluctuations.  Owing to this quantum analogy, an Efimov-DNA could be
an affordable system in the domain of classical biology for studying
aspects of the quantum Efimov physics.  In this paper we widen the
scope of the Efimov physics by establishing the presence of the effect
in certain classes of low-dimensional DNA models by staying purely in
the classical domain of statistical mechanics.  We also show that the
same cause that produces the Efimov-like effect in DNA can lead to a
new phase in triple-stranded DNA, a phase we call a {\it mixed phase}.

In 1970, a novel phenomena in quantum mechanics, the Efimov
effect \cite{efimov1,efimov2}, was discovered, which resembled the
by-then-forgotten Thomas effect of 1930's \cite{thomas}.  Three nucleons
with a critical short range pair potential become bound due to an
emergence of a long range interaction.  The result was a tower of an
infinite number of bound states right at the critical threshold of the
two-body binding.  As one moves away from the critical point the
number of bound states decreases and vanishes at a particular
strength.  This three-body bound state has a size much larger than the
range of the short range pair potential.  Such a loose three-body
bound state is named as the quantum Efimov state.

The paths of particles in quantum mechanics (QM), in the path integral
formalism, are analogous to Gaussian polymers under an imaginary time
transformation; the time of quantum mechanics maps on to the contour
length of the polymers.  In QM, along the paths of two interacting
particles, the interactions are strictly at the same time only.  This
maps nicely onto the sequential base pairing of a dsDNA.  The
excursions of the quantum particles in the classically forbidden
region because of quantum fluctuations correspond to the bubbles on
a DNA generated by thermal fluctuations.  The infinite time limit in
QM corresponds to an infinitely long DNA, a necessity for a phase
transition.  For the case of base pairing as the only form of mutual
interaction, the melting is equivalent to the unbinding transition of
a pair of particles in quantum mechanics when the bound state energy
approaches zero by tuning the potential.  This basic connection
prompts the similarities between the Efimov problem in QM and a tsDNA.

Triple-stranded DNA (tsDNA) is well-known in
biology \cite{rich57,buske}.  The base sequence of a double stranded
DNA (dsDNA) allows a third strand to bind via the Hoogsteen or the
reverse Hoogsteen pairing to form a triple helix \cite{frank95,radha}.
There are evidences, from NMR, of Hoogsteen pairing formed dynamically
(1\% of time) even in a normal DNA \cite{JACS2012}. The triplex helix
can also be formed with DNA-RNA \cite{Roberts} and DNA-peptide nucleic
acid (PNA), whose uncharged peptide backbone helps in the
stabilization of the triplet
structure \cite{Nielsen,ray00,sun96,Betts}.  A triple helix formation
controls the gene expression, which may be of use in
antibiotics \cite{antibio}, and therapeutic applications like targeting
a specific sequence in gene therapy \cite{therapy1,therapy2,helen}.
All of these involve tightly bound states of a size determined by the
hydrogen bond length.  The Efimov-DNA however is not a tight bound
state like these triple helices and one does not need any special
pairing for its formation.

The nature of dsDNA melting depends on many
factors and could either be an all-or-none process or be mediated by
the formation of bubbles along the chain.  Bubble formations increase
the entropy of the bound state.  The melting takes place when the gain
in entropy by strand unbinding outweighs the energy gain of the bound
state.  At or close to the duplex melting, if a double stranded DNA
allows bubbles of any length, a third strand of DNA can pair with the
strands of the bubble.  This process, known in biology as the strand
exchange mechanism, would lead to a bound state of the three together.
The possibility of a long-range attraction, an important aspect of the
Efimov effect, has been argued by a polymeric scaling analysis in
Ref.~\cite{efidna}.  The existence of a three-strand bound state has
further been verified by real space renormalization group (RG) on
hierarchical lattices of dimensions $d>2$, transfer matrix
calculations in real space in $1+1$ dimensions, and by an RG limit
cycle for polymer models in continuum in $3$
dimensions \cite{efidna,zeros,tanlim}.

\fignewx

The triplex formed by the pairwise attraction of bases has a melting
point higher than the duplex melting temperature.  As already
mentioned, the Efimov-DNA occurs in the region between the melting
point of dsDNA and tsDNA.  This is an exotic state mainly because of
the special role played by the third strand, but thermodynamically it
is not a distinct phase.  It is a continuation of the low temperature
triplex bound state.  This raises an interesting issue whether the
Efimov like state mediated by the third strand of DNA can be
stabilized as a thermodynamic phase, distinct from the triplex and the
denatured state.  This mixed phase, alluded to at the beginning, is a
bound state where, in any stretch of length, one strand remains
unbound with two others paired; it should share a boundary with the
denatured DNA on the high temperature side and a boundary with the
triplex state on the low temperature side.  We establish in this paper
that such a mixed phase {\it does occur} if the bubble formation on
the DNA is controlled suitably.  The major consequence of this
intermediate mixed phase is that a tsDNA would undergo two phase
transitions, triplex$\leftrightarrow$mixed$\leftrightarrow$denatured,
as opposed to a simple melting.  See Fig.~ref{fig:newx}.  This is one
of the important results of this paper.

It may now be asked, what it is that is responsible for the Efimov
effect.  For a broader perspective, it helps to define the DNA melting
problem in any dimension, like in many other polymer problems.  On one
hand, the standard quantum mechanical results and the polymeric
scaling argument indicate the importance of large scale fluctuations
in bubbles to produce an effective inverse square law
attraction \cite{efimov1,efidna}.  On the other hand, the models of DNA
on hierarchical lattices ($d>2$), which do not have any metric, also
show the Efimov-DNA.  It is then tempting to hypothesize that {\it the 
Efimov effect of three being bound but no two, is a consequence of a 
phase transition through its associated non-analytic behaviour}.  If
true, this would broaden the range of situations where the Efimov
effect could be seen.  Admittedly, it is difficult to establish the
hypothesis in the quantum domain but it can be done in the DNA
context.  For example, in lower dimensions ($d\le 2$), the bubble
entropy is not enough to cause a melting, so that DNA would remain
bound at all temperatures for any arbitrarily weak short-range
attraction \cite{comment}.  However, for a DNA in a lower dimension
$d\le2$, a phase transition can be triggered by adding extra factors.
These extra factors are either local constraints in bubble opening 
(e.g. crossing) or the hard core repulsion or some
cooperativity weight factors ($\sigma$) for each bubble formed in the
model between the DNA strands.  A test of the hypothesis would then be
to show the existence of an Efimov-DNA in such low dimensional models
with phase transitions.

In this paper we aim to verify the robustness of such a finding and to
reach such a purpose we want to investigate the effects of variation
of the $\sigma$ parameter. Hence it is worthwhile to consider a model
in a lower dimensional lattice which is amenable to exact treatments
and where $\sigma$ the cooperative parameter can be easily tuned. This
lattice turns out to be the Sierpinski gasket, the common regular
fractal which have been widely used in studying different statistical
models, for instance the Ising model, the directed or the self
avoiding polymer model, the Potts model, the sandpile model, the
ice-type vertex models, models of polymers under a force 
\cite{dhar,sgflavio,yash,critpheno,knez,elez,ice}.  It is remarkable
that very often the results obtained on Sierpinski gasket or on other
hierarchical lattices turn out to be true in the real world, for
instance, the collapse transition ($\theta$) point for linear
polymer \cite{dharvan,OSST91} or, as we will show later in the paper,
the correct order (first) of the denaturation transition of 
double-stranded DNA.

\section{Outline}
This paper is organized as follows.  In Sec.~\ref{ketmodel}, our model
on a Sierpinski gasket is introduced.  In Secs.~\ref{CNtwoc} and 
\ref{CNthreec} two- and three-polymer problems on a fractal
lattice are introduced.  The exact recursion relations of the
partition functions for both the crossing and the noncrossing cases
are written and the method of calculations is discussed.  The
two-chain phase diagrams are discussed.  With various interactions and
the crossing or the noncrossing conditions three different models of
the three-chain system are introduced.  Results obtained from the
exact recursion relations are discussed in Secs.~\ref{endis} and 
\ref{kettable}.

\section{model}\label{ketmodel}
A Sierpinski gasket is a fractal lattice obtained after an infinite
iteration from a single equilateral triangular lattice.  This
particular lattice is drawn on the two dimensional ($d=2$) plane.
Taking out the middle piece of a triangle yields three smaller
triangles and, by repeating this for every allowed triangle, the fractal
lattice is formed recursively; see Fig.~\ref{fig:gsmodel}.  The
dimension for an infinite lattice is
\begin{equation}
d=\frac{\ln N_n}{\ln L_n}=\frac{\ln 3}{\ln 2}\approx 1.58,
\end{equation}
where $N_n$ is the number of the surviving triangles and $L_n$ is
the number of bonds of the lattice along any one side of the lattice
at the $n$th generation.  

\GSmodel

In order to mimic the Poland-Scheraga \cite{PS66} DNA-like models in
which monomers in different strands interact only if their position
along the chain is the same (complementary bases), we consider
directed polymers on a Sierpinski gasket which are restricted to
occupy only the nonhorizontal bonds as shown in Fig.~\ref{fig:gswalk1}.  
In such a way, each time two different strands occupy the same bond, 
it is automatically guaranteed that they share the same chemical 
distance from the origin.  Still, there can be
two different classes of models differing in the restrictions on the
crossing of the two strands.  Two configurations with a bubble are
shown for generation $n=1$ in Figs.~\ref{fig:gswalk1}(a) and 
\ref{fig:gswalk1}(b).  If we allow crossing, the strands can exchange 
and both Figs.~\ref{fig:gswalk1}(a) and \ref{fig:gswalk1}(b) are allowed.  
In the noncrossing case, only (a) is allowed.  The crossing among the 
polymers increases the number of configurations, resulting in more entropic 
contributions compared to the noncrossing case.

\GSwalkM
In this approach the sequence of bases is not explicitly considered
since the model is coarse grained in character. In this respect each
monomer is not to be thought of as a single base, but as a group of
bases (block). Consequently a mismatch between corresponding blocks
has to be very disfavored with respect to a correct matching. 

We can consider two or three different polymers.  The following weights 
are assigned to them: 
\begin{itemize}
\item[$\bullet$] Fugacity $z$ for each bond,
\item[$\bullet$] Boltzmann factor $y_{ij}=e^{\beta\epsilon_{ij}}$,
  when a single bond is shared by the two polymers $i$ and $j$ with
  binding energy $\epsilon_{ij}$,  and $y_{ijk}=e^{\beta\epsilon_{ijk}}$
  when a single bond is shared by the three polymers with the binding
  energy $\epsilon_{ijk}$.  Here $\beta$ represents the inverse
  temperature $T$, $\beta=1/k_{\rm B} T$, where $k_{\rm B}$ is the
  Boltzmann constant.
\item[$\bullet$] $\sigma_{ij}$ for the two-chain and $\sigma_{ijk}$
  for the three-chain bubble opening or closure.
\end{itemize}
The weight of a walk of a single chain of length $N$ is $z^N$, where
$N$ is the number of bonds.  Usually \cite{dhar,OSST91,dharsingh} one can
consider $z$ as an extra variable, the fugacity for the length of the
polymers in a grand-canonical ensemble, but here we will set it to
$1$, as is discussed below. We use $z$ when a direct computation of
the free energy is required.

There are two special values of $\sigma$; $\sigma=1$ implies that no
weight is given for bubble opening or closure, and $\sigma=0$ implies
no bubble formation, {\it i. e.}, a model without any bubble (fork
model).  In biological contexts the co-operativity factors $\sigma$'s
depend, for example, on the chain length, the ionic strength, the stacking
potential, etc. \cite{ionic}.  Most of the studies have reported the
value of cooperativity factor in the range $10^{-4}$--$10^{-5}$ with
the loop nucleation free energy as $\sim -k_BT\ln\sigma\sim10k_BT$.  
We shall take the cooperativity factor as a controlling parameter, not
necessarily restricted to small values.

To study the melting of DNA on a fractal lattice, we need to define
the partition functions for the two- and the three-chain systems as
shown in Fig.~\ref{fig:gswalk}.  We choose $z=1$ to be in the
canonical ensemble.  The standard way to study the polymers on a
fractal lattice is to find out the fixed point of $z$ by an RG
procedure as proposed by Dhar \cite{dhar}.  This corresponds to the
grand canonical ensemble, where the fixed point of $z$ gives the free
energy.  We know that the choice of ensemble does not matter, as long
as we work with the large lengths of the polymers.  In our approach we
calculate the free energies of different possible phases in the 
canonical ensemble, look for the most favorable one, and obtain the
phase diagram directly from the free energies.  Since all the polymers
are of same length ($N=2^{n+1}\to\infty$) and traverse the whole
lattice, we may set $z=1$.

Different possible polymer configurations are shown in
Fig.~\ref{fig:gswalk}.  The corresponding partition functions $a_n$,
$b_n$, $c_n$, $d_n$, $e_n$, $f_n$, $g_n$, $h_n$, $i_n$ are defined at
the $n$th generation and their corresponding recursion
relations can be easily computed for successive generations.  An
example of the procedure to obtain the recursion relation is given in
Appendix A.
The initial conditions (the partition functions at the first stage of 
iteration) are dictated from the physical properties of the studied model.

\GSwalk

\section{Two stranded DNA on the Gasket}\label{CNtwoc} 
In order to explain our strategy and to fix some preliminary results
let us first consider the melting of a double stranded DNA.  The
partition functions of a single chain and a double chain for the $n$th
generation are given by $b_n$ and $d_n$ respectively. However, to do
the sum over all configurations, one needs the sub-partition
functions, $a_n, c_n$ and $g_n$ as one sees from Fig.~\ref{fig:gswalk1} 
and Appendix A.  The crossing and the noncrossing cases are discussed 
separately below.

\subsection{With crossing}
We first consider a two-chain system where the walks can cross each
other.  Here $y$ is the weight at the bond for sharing it by the two
polymers.  The two-chain bubble opening or closure is associated with
the weight $\sigma$ at the vertex.  Five partition functions are
necessary and using the label used in Fig.~\ref{fig:gswalk} their
values for the $(n+1)$th generation are given by (see Appendix
A)
\begin{subequations}
\begin{align}
a_{n+1}&=a_n^2,\label{p1}\\
b_{n+1}&=b_n^2+a_n^2b_n,\label{p2}\\
c_{n+1}&=c_n^2,\label{p3}\\
d_{n+1}&=d_n^2+2g_n^2b_n+c_n^2d_n,\label{p5}\\
g_{n+1}&=a_ng_n(b_n+c_n).\label{p4}
\end{align}
\end{subequations}
The Boltzmann factors and other weights are defined on the bonds and 
the sites.  They are therefore specified for the smallest triangle, 
i.e., at the zeroth generation.  Those specifications act as the 
initial conditions for the recursion relations.
The initial conditions are taken as 
\begin{equation}\label{in1}
a_0=1,~ b_0=1,~ c_0=y,~ d_0=y^2,~ g_0=y\sigma.
\end{equation}
These values follow from Fig.~\ref{fig:gswalk} by counting the shared
  bonds and bubble opening or closing.  For $c_0$ and $g_0$ there is
  only one bond with two  strands on it
  and hence they require a Boltzmann factor $y$.  On the other hand, $d_n$ 
  has two shared
  bonds, thereby requiring a factor $y^2$.   A configuration like $g_n$ is
  required to open or close a bubble.  Hence $g_n$ involves an
  additional $\sigma$ for the junction point.

By iterating the equations it turns out that the leading terms are
coming from the generating function $b_n$ (single chain) and $d_n$
(two chains).  It is then convenient to look at the ratio
\begin{equation}\label{cond1}
r_1=\frac{d_{n+1}}{b_{n+1}^2}.
\end{equation} 
\cmodel  
This ratio compares the two-chain bound state free energy with the
free energy when the two strands are in the denatured state.  By
monitoring the divergence or the convergence of $r_1$, for given
values of $\sigma$ and $y$, one can easily and quickly pinpoint the
denaturation transition and obtain the phase diagram in the
$y$-$\sigma$ plane.  The phase diagram is shown in Fig.~\ref{fig:cross}.  
The transition is from the unbound to the bound state of the 
two-stranded DNA at $y=y_c(\sigma)$.

For $\sigma=0$, there are no bubbles. In this situation, the bound
state partition function for the $n$th generation consists of two
factors, the Boltzmann factor and the number of configurations of 
the bound pair $b_n$, i.e.
\begin{equation}
  \label{eq:3}
  d_n= b_n \; y^N,\quad N=2^{n+1},
\end{equation}
while the partition function of the unbound state is $b_n^2$.  The
continuity of the free energy at the transition point ($r_1=1$) then
gives the transition point as
\begin{eqnarray}
\label{eq:yczero}
\ln y_c(0) &=& \lim_{n\to\infty} \frac{1}{2^{n+1}}\ \ln b_n, \nonumber\\
{\rm or,}\quad y_c(0)&=&1.2640847353... .
\end{eqnarray}
This value corresponds to the temperature at which the binding energy
per bond $\epsilon$ is equal to the entropic free energy ($T s$) of a
single chain, viz., 
\begin{equation}
  \label{eq:4}
  T_c= \epsilon/s,\quad {\rm with}\ s=\lim_{n\to\infty} (1/N)\ln b_n.
\end{equation}

\subsection{No crossing}
If the crossing between the two strands DNA is not allowed, the
recursion relations are same as the crossing case except for $d_n$,
which in this case is
\begin{equation}
d_{n+1}=d_n^2+g_n^2b_n+c_n^2d_n. \label{eq:2}
\end{equation}
The initial conditions are still given by Eq.~(\ref{in1}).  A similar
comparison method [Eq.~(\ref{cond1})] is used here as in the two-chain
crossing case.  We obtain the phase diagram in the $y$-$\sigma$ plane
as shown in Fig.~\ref{fig:cross}(b).  For $\sigma=0$ the two-chain
melting is at $y_c(0)=1.264...$, which is the same as in the crossing
case.  There is a difference between the crossing and the noncrossing
melting curve for $\sigma\ne 0$.  In fact, the two curves can be
mapped onto one another by rescaling $\sigma$ by $1/\sqrt{2}$ in the
crossing case; indeed, the consequent rescaling of $g_n$ necessary to
keep the initial condition in the form of (\ref{in1}) allows us to
change Eq.~(\ref{eq:2}) into Eq.~(\ref{p5}). In particular it can be
noticed that for $\sigma=1$ the melting transition occurs at a finite
temperature only for the noncrossing model.  For the crossing case,
$y_c(\sigma=1)=1$, but for the noncrossing case
$y_c(\sigma=\sqrt{2})=1$.

It is important to notice that for both the considered models (with
crossing and no crossing) and for any $\sigma$ , the first derivative
of the free energy is discontinuous at the thermal transition 
(see Fig.~\ref{fig:ss2}).  Therefore, despite its simplicity, our model 
predicts a first order transition for DNA denaturation as observed
experimentally \cite{Kafri}.

\section{Three strands}\label{CNthreec}
When we consider the three-chain system, several cases are possible.
With the crossing and the noncrossing conditions and the choices of
the interacting and the noninteracting pairs, we classify different
models.  Among the many possible varieties we will discuss only three 
of them, TS1, TS2, and TS3 since they exhibit the full range of critical
behaviours we explored. The models are the following:
\begin{enumerate}
\item Model TS1: This is the noncrossing case with a weight for
  two-chain bubble opening or closure of all pairs. The weight is
  penalizing bubbles for $\sigma<1$ but favoring for $\sigma>1$.
  There is no contact energy between chains $1$ and $3$.  These two
  chains are nevertheless coupled with each other through the $\sigma$
  weight, only when all the three strands are bound together.  As a
  consequence, the opening or closure in the triplex state is weighted
  twice (it involves two pairs) with respect to the duplex state.
\item Model TS2: This is the crossing case with the three-chain
  repulsion, so that the overall energy of the triplex state is the
  same as for the duplex state.  Similar to TS1, the weight for a
  two-chain bubble opening or closure is present for all pairs, so
  that the opening or the closure in the triplex state is weighted
  twice with respect to the duplex state.
\item Model TS3: This is the crossing case with the three-chain
  repulsion and a weight for both the two- (for all pairs) and the
  three-chain bubble opening or closure.  The weight for the
  three-chain bubbles counters that for two-chain ones, so that both
  the overall energy and the weight for the opening or closure are the
  same in the triplex and in the duplex state.
\end{enumerate}

\subsection{Model TS1: Noncrossing}   
In this case walks can not cross each other.  We assign a weight
Boltzmann factor $y$ for each interaction between chains 1 and 2, and
2 and 3 {\it i. e.}, $y_{12}=y_{23}=y$, but no interaction between
chains 1 and 3, {\it i. e.}, $y_{31}=1$.  The weight $\sigma$ is
assigned for each bubble opening between all pairs, {\it i. e.},
$\sigma_{12}=\sigma_{23}=\sigma_{31}=\sigma$.  When all chains are
together we consider a weight $y^2$ and such a situation can also be
described if we take $y_{12}=y_{23}=y_{31}=y$ and $y_{ijk}=1/y$.  If
$y>1$, $y_{ijk}$ is repulsive in nature.  The two definitions of the
contact energies are equivalent only because of the noncrossing
constraint.

\comodel 

The recursion relations for the partition functions for this model are given by 
\begin{subequations}
\begin{align}
a_{n+1}&=a_n^2,\label{D1e1}\\
b_{n+1}&=b_n^2+a_n^2b_n,\\
c_{n+1}&=c_n^2,\\
d_{n+1}&=d_n^2+g_n^2b_n+c_n^2d_n,\\
e_{n+1}&=e_n^2,\\
f_{n+1}&=f_n^2+e_n^2f_n+h_n^2d_n+i_n^2b_n,\label{D1e6}\\
g_{n+1}&=a_ng_n(b_n+c_n),\\
h_{n+1}&=h_n(a_ne_n+b_nc_n),\\
i_{n+1}&=i_n(c_ne_n+d_na_n)+g_n^2h_n,\label{D1e9}
\end{align}
\end{subequations}
with initial conditions:
\begin{eqnarray}
&a_0=1, b_0=1, c_0=y, d_0=y^2, e_0=y^2, f_0=y^4, \nonumber \\
&g_0=y\sigma, h_0=y^2\sigma^2, i_0=y^3\sigma^2.
\end{eqnarray}
The powers of $y$ follow from Fig.~\ref{fig:gswalk} by counting the
pairs sharing the bonds.  For the $\sigma$ factors, we note that both
$h_0$ and $i_0$ correspond to a single chain breaking off from a
triplet, thereby producing two ``bubbles'' with the remaining two.
Hence $\sigma^2$ for both these partition functions.  For example, in
$i_0$, one bond with three chains has three pairs requiring $y^3$ but
with an additional factor $y_{123}=1/y$ for the three chain
interaction, while the other bond has only one pair requiring a
factor $y$.   This gives $y^3$ with $\sigma^2$ for opening or closing
of two bubbles.

We look at the divergence or convergence of the ratios 
\begin{eqnarray}
r_2&=&\frac{f_{n+1}}{b_{n+1}^3},\label{rat2}\\ 
r_3&=&\frac{f_{n+1}}{b_{n+1}d_{n+1}}\label{rat3}, 
\end{eqnarray}
for given $\sigma$ and $y$.  The idea behind the choice
of the above  ratios is to compare the three-chain free energy with the
free energy when three chains are free [$r_2$ in Eq.~(\ref{rat2})],
or when one chain remains isolated with the other two forming a duplex
[$r_3$ in Eq.~(\ref{rat3})].

By looking at the divergence or convergence of the ratios $r_2$ and
$r_3$ for different values of $y, ~\sigma$ and comparing these values
with the two-chain melting curve, different phases can be identified 
(see Fig.~\ref{fig:s1}).  In this model we obtain two different phases,
an Efimov and a mixed phase.  However the Efimov phase is not a
distinct phase.  It is just an effect on three chains, where no two
are bound but three are bound.  On the other hand, in a mixed phase,
the strands are pair-wise bound but no three-chain contact.  The
possible types of the mixed phase are shown schematically in 
Fig.~\ref{fig:mixed}.  In Fig.~\ref{fig:s1}, within the range $y=1$ to
$y<1.07526$ for $\sigma >1.14458$ the Efimov region is obtained and
the region is enclosed between the line for $r_2$ and the two-chain
melting curve.  The mixed phase is enclosed between the line for $r_3$
and the two-chain melting curve for $y>1.07526$ and $\sigma
<1.14458$.  Unlike the Efimov DNA, the mixed phase undergoes a phase
transition to a state of three-chain bound state.

 \mixed

\subsection{Model TS2: With crossing}
We now extend the study to a slightly different model with the following 
characteristics:
\begin{itemize}
\item Walks can cross each other.
\item $y_{12}=y_{23}=y_{31}=y,~ y_{123}=\frac{1}{y}$.
\item {$\sigma_{12}=\sigma_{23}=\sigma_{31}=\sigma,~ \sigma_{123}=1$.}
\end{itemize}
In this model all chains are having equal pair interaction.  There is
a three-chain repulsive interaction.  A weight is given for the
two-chain bubble opening or closure for all pairs.  
With crossing,
there will be extra weights for configurations involving bubble
opening due to the  exchange of strands.  Therefore, configurations
involving $g_n,h_n$, and $i_n$ would have additional combinatorial
factors compared to model TS1 (Eqs.~\eqref{D1e1}-\eqref{D1e9}).  
The recursion relations for the ${(n+1)}$th generation partition 
functions are given by
\begin{subequations}
\begin{align}
a_{n+1}&=a_n^2,\label{D2e1}\\
b_{n+1}&=b_n^2+a_n^2b_n,\\
c_{n+1}&=c_n^2,\\
d_{n+1}&=d_n^2+2g_n^2b_n+c_n^2d_n,\\
e_{n+1}&=e_n^2\\
f_{n+1}&=f_n^2+e_n^2f_n+3h_n^2d_n+3i_n^2b_n,\label{D2e6}\\
g_{n+1}&=a_ng_n(b_n+c_n),\\
h_{n+1}&=h_n(a_ne_n+b_nc_n),\\
i_{n+1}&=i_n(c_ne_n+d_na_n)+2g_n^2h_n\label{D2e9},
\end{align}
\end{subequations}
with the initial conditions
\begin{eqnarray}
&a_0=1, b_0=1, c_0=y, d_0=y^2, e_0=y^2, f_0=y^4,\nonumber\\
&g_0=y\sigma, h_0=y^2\sigma^2, i_0=y^3\sigma^2.
\end{eqnarray}
\aomodeltwo
Following the same procedure of comparison of free energies, the phase
diagram is obtained in the $y$-$\sigma$ plane, as shown in
Fig.~\ref{fig:s22}.  With the given initial conditions this model
exhibits the mixed phase.  One sees two transitions: At low
temperature we have a three-chain bound state that goes into the mixed
state (green dash-dotted line in Fig.~\ref{fig:s22}) and the mixed state
melts into free chains (black solid line in Fig.~\ref{fig:s22}).  This
latter transition coincides with the two-chain melting curve.

\subsection{Model TS3: With crossing}
Here three chains have repulsive interaction as in TS2, but we
consider a different generalization that favours three-chain bubbles.
\begin{itemize}
\item Walks can cross each other.
\item $y_{12}=y_{23}=y_{31}=y,~ y_{123}=\frac{1}{y}$.
\item {$\sigma_{12}=\sigma_{23}=\sigma_{31}=\sigma,~ 
\sigma_{123}=\frac{1}{\sigma}$.}
\end{itemize}
Here $\sigma<1$ and therefore $\sigma_{123}>1$.  Two-chain bubbles are penalized 
by $\sigma$ but $\sigma_{123}$ favours three-chain bubbles.   

\aomodelthree

However the recursion relations are same as for TS2 given by
Eqs.~(\ref{D2e1})-(\ref{D2e9}). The initial conditions are
\begin{eqnarray}
&a_0=1, b_0=1, c_0=y, d_0=y^2, e_0=y^2,&\nonumber\\
&f_0=y^4, g_0=y\sigma, h_0=y^2\sigma, i_0=y^3\sigma.&
\end{eqnarray}
Following the same procedure of comparison of free energies, the Efimov 
state is obtained and is shown in Fig.~\ref{fig:s23}.  

\section{Energy diagram}\label{endis}
The first order nature of the phase transitions can be determined from
the behaviour of the average energy.  For that we first validate, with a
direct calculation of the free energy, the identifications of
the phases done in the previous sections.

\fugac

In the grand canonical approach, we determine its fixed point value of
fugacity $z$ (See Sec.~\ref{ketmodel}), for given values of $y$ and
$\sigma$. These are shown in Fig.~\ref{fig:compar}.  Based on the idea
of the various phases, the total partition functions $Z_{\rm tot}$ and
$Q_{\rm tot}$ for the two-chain and the three chain cases in the fixed
length ensemble can be written as
\begin{eqnarray}
  Z_{\rm tot}&=&b_{n+1}^2+d_{n+1},\label{ECgasket1}\\
  Q_{\rm tot}&=&f_{n+1} + b_{n+1}^3+2d_{n+1}b_{n+1},\label{ECgasket2}
\end{eqnarray} 
in terms of the subpartition functions $b_{n+1}$, $d_{n+1}$, and
$f_{n+1}$.  A comparison of the grand canonical partition function $z$
and the canonical one $Q_{\rm tot}^{1/3N}$ for polymers of length
$N=2^{n+1}$ with $n=25$, is shown in Fig.~\ref{fig:compar}.  
The figure also shows the partition function for the mixed phase 
$[b_n d_{n}]^{1/3N}$ vs $y$.  Armed with this agreement, the
average energy calculation can be simplified.  The total average
energies of the two-chain system ($E_{\rm tot}$) and the three-chain
system (${\cal E}_{\rm tot}$) can be written as
\begin{eqnarray}
E_{\rm tot}&=&\frac{d_nE_{d_n}}{Z_{tot}},\\
{\cal E}_{\rm tot}&=&\frac{f_nE_{f_n}+2b_nd_nE_{d_n}}{Q_{\rm tot}}
\end{eqnarray} 
where $E_{d_n}$ and $E_{f_n}$ are the energies corresponding to the
partition functions $f_n$ and $d_n$, all of which can be computed
iteratively.  The recursion relations for the energies for model TS1 
are given in Appendix B.

The three-chain average energy per bond, 
$\langle E\rangle= {\cal E}_{\rm tot}/N$, is shown for model TS1 in
Fig.~\ref{fig:ss2}.  Fig.~\ref{fig:ss2}(a) is for $\sigma=1.25$.  The
three-chain average energy (marked as 1) is compared to the two-chain
average energy (marked as 2).  This shows the nonzero three-chain
average energy, even though the duplex average energy is zero. 

\senergy

Fig.~\ref{fig:ss2}(b) is for $\sigma=0.5$.  The three-chain average 
energy (marked as 1) is compared to the two-chain average energy 
(marked as 2).  The transition from the unbound to the mixed state 
is at the same temperature as the two-chain case, {\it i. e.}, 
at $y_c(\sigma)$.  The transition from the mixed state to the bound 
state occurs for $y>y_c(\sigma)$ (lower temperature).   

The average energy curve in Fig.~\ref{fig:ss2}(a) marked as 1 shows only one 
jump, where as in Fig.~\ref{fig:ss2}(b) the average energy curve marked as 1 
shows two jumps.  In the latter case the two transitions are from the unbound 
to the mixed state and from the mixed to the three-chain bound state.

\section{Discussion}\label{kettable}

All the models and results are given below for easy reference.  A
large class of models can be defined distinguished by the nature of
interactions and the cooperativity factors.  Many of the models are
not discussed in details in this paper but the results are stated in
Table.~1.

We discuss briefly model TSnull, because it is a reference model that
allows us to understand the origin of the three-body effects, either
Efimov or the mixed state, in the other models. In model TSnull,
instead, the duplex and the triplex melting curves superpose exactly
and no special three-body effect is present. That is due to chains $1$
and $3$ being uncoupled, so that the three-chain behaviour is dictated by
the independent behaviour of the chain pairs $12$ and $13$.  

Any model feature that effectively couples chains $1$ and $3$ causes
the presence of cooperative three-body effects. The coupling can be
induced by conditions on the contact energies $y (>1)$'s, on the weights
$\sigma$'s for bubble opening and closure, or by the presence of the
noncrossing constraint.  Depending on the combination of those
conditions, a few models, like TS2, show the mixed state while
the others, like TS3, show the Efimov-like state.  But for model TS1 
we get both of the states though in different regimes of $\sigma$ and $y$.

If we compare models TS2 and TS3, in both of them the overall energy
of the triplex state is the same as for the duplex state due to the
repulsive nature of the three-chain interaction ($y_{123}>1$), but
there is a bias in TS2 penalizing the bubble opening or closure in the
triplex state. This biasing seems to favor the mixed state in TS2, by
entropically destabilizing the triplex state.  On the other hand, the
conditions on the $\sigma$'s used in TS3 remove this bias and leave an
effective coupling between chains $1$ and $3$ that seems to favour the
Efimov state by entropically stabilizing the triplex state.
Intriguingly, the Efimov state is stabilized through the same
mechanism in model TS4 as well, even in the absence of the energetic
coupling between chains $1$ and $3$ that is present in both models TS2
and TS3.

The presence of the noncrossing constraint further complicates
things.  Its effect on a two-chain system is equivalent to a rescaling
of $\sigma$ by a $1/\sqrt{2}$ factor in the presence of crossing, thus
causing the entropic destabilization of the duplex state.  In a three-chain
system a different rescaling by a $1/\sqrt{3}$ factor would be needed
to obtain Eq.~(\ref{D1e6}) from the corresponding Eq.~(\ref{D2e6}) in
the presence of crossing.  As a consequence, the simultaneous presence
of two-chain and three-chain bubbles does not allow to establish any simple
mapping between the noncrossing model and a $\sigma$-rescaled
crossing model.  Yet, one can argue on this basis that the noncrossing
constraint induces an entropic destabilization stronger for the
triplex state than for the duplex.  In fact, in model TS5 the coupling
between chains $1$ and $3$ is due only to the noncrossing constraint,
and the mixed phase emerges, consistent with the above observation.

Finally, in model TS1 a further coupling is caused by the choice of
the $\sigma$'s weights that either penalizes (for $\sigma<1$) or
favours (for $\sigma>1$) the bubble opening or closure in the triplex
state. As a result, the mixed and the Efimov states coexist in the
same phase diagram, with the Efimov state being present in the
$\sigma>1.14458$ part of the phase diagram.

For $\sigma=0$ all the models are like the Y-fork model and come out
to be the same, and $y_c(0)=1.2640847353...$ denotes the melting for
both the two- and the three-chain systems.

All the models that we considered show first order phase transitions,
with discontinuities in the average energy.  This is an effect due to
the fractal lattice since similar DNA models defined through directed
polymers on the Euclidean lattice show second order melting
transitions when $\sigma>0$ \cite{unzip2}.  Only for the Y-fork model
the melting transition is first-order on the Euclidean lattice as
well.

\newcommand{\tablone}{%
\begin{center}
\begin{table}[htbp]
\centering
\begin{tabular*}{.85\textwidth}{@{\extracolsep{\fill}} | c | c | c | r | }
  \hline
Model&Parameters &Parameters& Results\\
  \hline
TS1 (Noncrossing)& $y_{12}=y_{23}=y,y_{31}=1$ & $\sigma_{ij}=\sigma$,~$\sigma_{123}=1$& Efimov, Mixed\\
\hline
TS2 (Crossing)& $y_{ij}=y, y_{123}=1/y$& $\sigma_{ij}=\sigma,\sigma_{123}=1$ &Mixed\\
\hline
TS3 (Crossing)& $y_{ij}=y, y_{123}=1/y$ & $\sigma_{ij}=\sigma, \sigma_{123}=1/\sigma$&Efimov\\
 \hline
TS4 (Crossing)& $y_{12}=y_{23}=y, y_{31}=1$ &$\sigma_{ij}=\sigma$,~$\sigma_{123}=1/\sigma$ &Efimov\\
 \hline
TS5 (Noncrossing)& $y_{12}=y_{23}=y,y_{31}=1$ & $\sigma_{ij}=\sigma$,~$\sigma_{123}=1/\sigma$ & Mixed\\
\hline
TSnull (crossing)& $y_{12}=y_{23}=y,y_{31}=1$ & $\sigma_{12}=\sigma_{23}=\sigma, \sigma_{31}=1,\sigma_{123}=1$ & Nothing\\
\hline
\end{tabular*}
\caption{The results obtained for the three-chain models. The
  subscripts label the chains, $i,j=1,2,3$. The models are
  distinguished by the conditions satisfied by the parameters.  The
  new phases obtained are also flashed in this table.  Models TS4
  and TSnull require a different set of recursion relations, that
  use 14 generating functions, as shown in Appendix C.}
\end{table}
\end{center}
}

\section{Conclusion}
Working on regular fractal lattices has the advantage of allowing for exact
solutions.  Consequently, even very tiny and elusive effects, as those
observed in this paper, can be highlighted without the doubts that can
affect numerical simulations on Euclidean lattices.  For these
reasons, we believe our results are very intriguing and deserve
attention by experimentalists.

In particular we have shown that, when an extra weight $\sigma$ for
the two- and the three-chain bubble opening and closure is introduced
the Efimov-DNA, a loosely bound three-chain state where no two are
bound, occurs even in $d<2$. What is remarkable is the emergence of a
new state, to be called a mixed state, where locally any two are bound
keeping the third-strand always free but in a global view no one is
completely free.  The intermediate phase evolves as a separate phase
whereas the Efimov state is a crossover.

The cooperativity factor $\sigma$ acts as a control parameter for the
bubbles on DNA.  We see that for both Efimov-DNA and the mixed phase,
the existence of bubbles is a necessity. There is no such effect at
$\sigma=0$, despite a duplex melting transition. Since there is no
distance defined on the fractal lattices, the results of the paper do
not necessarily require any induced long range interaction. Our
results for a large varieties of models rather imply that a necessary
mathematical condition for both the phenomena is the bubble induced
thermodynamic phase transition.

Why DNA? The native interaction involving base pairs at the same
monomer position on the two strands is very special to DNA.  The
effects we are modeling depend crucially on this feature, even though
the strands can be taken as ordinary polymers.  For ordinary polymers,
monomers interact irrespective of their locations on the
chain \cite{SMB2013} which vitiates the quantum - polymer mapping, and
the models used here.  Fractal surfaces are routinely generated in the
laboratory but we are not aware of any attempt of adsorption of DNA or
any other polymers on such fractal objects. The closest we are aware
of is DNA adsorbed on a surface.  E.g. a double stranded DNA on a
lipid bilayer is known to behave like a two-dimensional self-avoiding
random polymer \cite{maier1999}.  We feel that such systems of DNA in
low dimensions might show some signature of the ``mixed phase''. We
tend to believe that DNA adsorbed on a surface is the most natural
choice for seeing the mixed phase predicted in this paper.

The existence of a bound state involving two otherwise denaturated 
strands of DNA due to the presence
of a third strand (the Efimov state) or the opening of a double for
the presence of a third strand, might have important implications for
biological processes.  Many biological processes involve three
strands, especially strand exchange.  Whether the emergent structures
resemble the phases obtained in this paper remains a matter of
speculation.  We expect our results will stimulate further theoretical
calculations in higher dimensions and new experiments to look for
signatures of the proposed mechanisms.

\begin{acknowledgments}
Support from Programmi di Ricerca Scientifica di Rilevante Interesse 
Nazionale is acknowledged by S.~M.~B. and F.~S.~ through Grant 
No. 2009SKNEWA, and A.~T.~ through Grant No. 2010HXAW77\_011.
\end{acknowledgments}

\begin{widetext}

\tablone

\appendix

\section{Diagrams for recursion relation}
In this appendix we show how to generate the recursion relations for
$d_n$ for the noncrossing case as an example. The sub-partition function for
$n+1$ can be expressed in terms of the various  partition
functions of the $n$th generation as shown in Fig.~\ref{fig:appx}. 
For the crossing case, one would need a factor of 2 for two
possibilities of the bubble  in Fig.~\ref{fig:gswalk1}.
\begin{figure*}[htbp]
\includegraphics[width=0.8\textwidth]{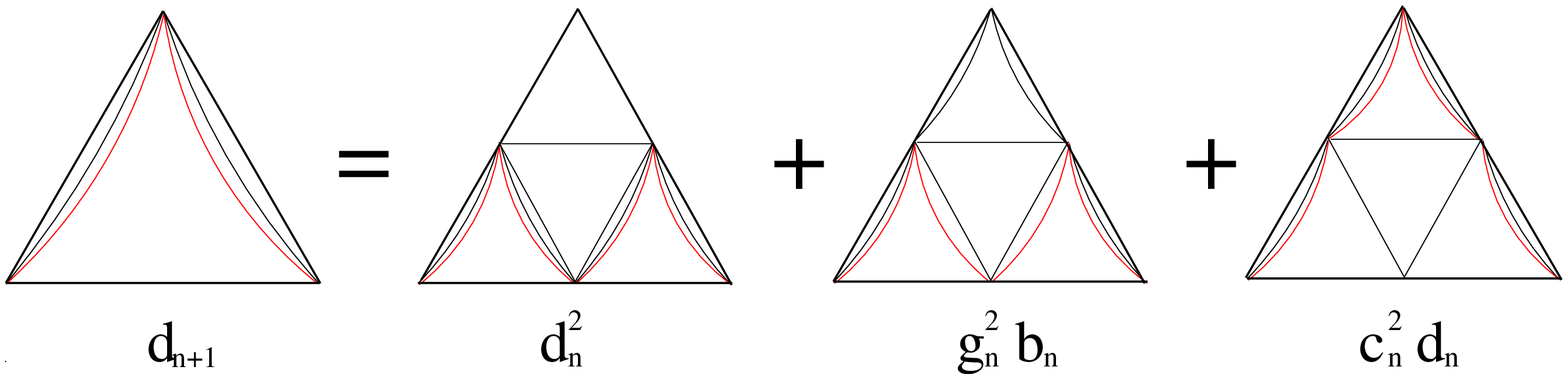}
\caption{Diagrams showing  Eq.~\eqref{eq:2}.  
A combinatorial factor 2 is needed for the second diagram on the 
right hand side for Eq.~\eqref{p5}, as explained in Fig.~\ref{fig:gswalk1}.
}
\label{fig:appx}
\end{figure*}

\section{Recursion relations for energies}
For model TS1, one may associate an energy for each of the sub-partition
functions.  These energies obey the following  recursion relations: 
\begin{eqnarray}
  \label{eq:1}
  E_{c_{n+1}}&=& 2 E_{c_n}\\
E_{d_{n+1}}&=&
\frac{1}{d_{n+1}} \left[2 d_n^2 E_{d_n} + c_n^2 d_n E_{d_n} + 2 c_n^2 d_n
           E_{c_n} + 2 g_n^2 b_n E_{g_n} \right]\\
E_{e_{n+1}}&=&2  E_{e_n}\\
E_{f_{n+1}}&=&\frac{1}{f_{n+1}}\left[2 f_n^2 E_{f_n} + e_n^2 f_n E_{f_n} + 2e_n^2 f_n E_{e_n}
              + h_n^2 d_n E_{d_n} + 2 d_n h_n^2 E_{h_n} + 2 b_n i_n^2 E_{i_n}\right]\\
E_{g_{n+1}}&=&\frac{1}{g_{n+1}}\left[(b_n + c_n) a_n g_n E_{g_n}
            + g_n a_n c_n E_{c_n}\right]\\
E_{h_{n+1}}&=&\frac{1}{h_{n+1}}\left[h_n E_{h_n} (a_n e_n + b_n c_n) +   h_n (a_n e_n E_{e_n}
            + b_n c_n E_{c_n} \right]\\
E_{i_{n+1}}&=&\frac{1}{i_{n+1}}\left[i_n E_{i_n} (c_n e_n + d_n a_n) + 
  i_n (c_n e_n E_{c_n}
             + c_n e_n E_{e_n} + d_n a_n E_{d_n}) + g_n^2 h_n E_{h_n} + 
  2 g_n^2 h_n E_{g_n}\right]
\end{eqnarray}
These are used to calculate the energies in Sec.~\ref{endis}.
\section{Recursion relations for TS4 and TSnull models}

In this appendix we show the recursion relations used for TS4 and
TSnull models.  In both cases we need to use an expanded set of 14
generating functions, because the conditions on the $y$'s and
$\sigma$'s parameters cause the chain pair $13$ to have different
properties with respect to the two other pairs $12$ and $23$.
Therefore, the two chain generating functions $c_n$, $d_n$, $g_n$ and
the two three-chain bubble opening/closure generating functions $h_n$,
$i_n$ needs to be considered twice. Note that if the noncrossing
constraint is present, conditions such as $y_{12}=y_{23}=y,y_{31}=1$
are equivalent to $y_{ij}=y, y_{123}=1/y$ and there is no need for an
extended set of generating functions.

\begin{eqnarray}
a_{n+1}&=&a_n^2,\\
b_{n+1}&=&b_n^2+a_n^2b_n,\\
c_{12,n+1}&=&c_{12,n}^2,\\
c_{13,n+1}&=&c_{13,n}^2,\\
d_{12,n+1}&=&d_{12,n}^2+2g_{12,n}^2b_n+c_{12,n}^2d_{12,n},\\
d_{13,n+1}&=&d_{13,n}^2+2g_{13,n}^2b_n+c_{13,n}^2d_{13,n},\\
e_{n+1}&=&e_n^2\\
f_{n+1}&=&f_n^2+e_n^2f_n+2h_{12,n}^2d_{12,n}+h_{13,n}^2d_{13,n}+2i_{12,n}^2b_n+i_{13,n}^2b_n,\\
g_{12,n+1}&=&a_ng_{12,n}(b_n+c_{12,n}),\\
g_{13,n+1}&=&a_ng_{13,n}(b_n+c_{13,n}),\\
h_{12,n+1}&=&h_{12,n}(a_ne_n+b_nc_{12,n}),\\
h_{13,n+1}&=&h_{13,n}(a_ne_n+b_nc_{13,n}),\\
i_{12,n+1}&=&i_{12,n}(c_{12,n}e_n+d_{12,n}a_n)+g_{12,n}^2h_{12,n}+g_{12,n}g_{13,n}h_{13,n},\\
i_{13,n+1}&=&i_{13,n}(c_{13,n}e_n+d_{13,n}a_n)+2g_{12,n}g_{13,n}h_{12,n}.
\end{eqnarray}

\end{widetext}



\begin{thebibliography}{99}
\bibitem{melt} The complete breaking of the base pairing hydrogen
   bonds of a dsDNA is called the melting of DNA.  It is also known as
   the unbinding transition, or the denaturation transition.  These names
   will be used interchangeably.  The two free strands in the ``molten''
   state are also called denatured DNA.  The force induced opening is
   called unzipping of DNA \cite{unzip1,unzip2,unzip3,kapri}.
\bibitem{comment} This is similar to the quantum mechanics result of
  at least one bound state in any attractive short range potential for
  $d\le 2$, while in higher dimensions the potential has to attain a
  critical strength to form a bound state.
\end{thebibliography}
\end{document}